\documentclass[aps,showpacs,prl,amsmath,floatfix,preprint]{revtex4}
\usepackage{graphicx,bm,epsfig}
\begin{document}


\def\beq{\begin{equation}}
\def\eeq{\end{equation}}
\def\bfg{\begin{figure}}
\def\efg{\end{figure}}
\def\cpt #1{ \parindent1in\parbox{4.5in}{\caption{\footnotesize #1}} }
\newcommand{\bra}[1]{\langle #1|}
\newcommand{\ket}[1]{|#1\rangle}
\newcommand{\braket}[2]{\langle #1|#2\rangle}

\title{Parametrically Shielding Electromagnetic Fields by Nonlinear Metamaterials}
\author{Simin Feng} 
\author{Klaus Halterman}
\affiliation{Physics and Computational Sciences  \\  Naval Air Warfare Center, China Lake, CA 93555}
\date{\today}

\begin{abstract}
An analytical theory is developed for parametric interactions in metamaterial multilayer structures with simultaneous nonlinear electronic and magnetic responses and with near-zero refractive-index.  We demonstrate theoretically that electromagnetic fields of certain frequencies can be parametrically shielded by a nonlinear left-handed material slab, where the permittivity and permeability are both negative.  The skin depth is tunable, and even in the absence of material absorption, can be much less than the wavelength of the electromagnetic field being shielded.  This exotic behavior is a consequence of the intricate nonlinear response in the left-handed materials and vanishing optical refractive-index at the pump frequency.
\end{abstract}

\maketitle


The recent advancement of fabrication technologies has opened up numerous avenues in designing metamaterials, which are composite structures with modified electromagnetic properties.  The main thrust of metamaterial design is to push the limits of ever increasing control over the effective material parameters, permittivity, $\epsilon$ and permeability, $\mu$.  By accessing previously forbidden regions of material parameter-space, many highly unusual electromagnetic properties and fascinating phenomena
have been predicted to occur, not to mention the possibility of novel device applications that may not be realized with conventional materials \cite{Smith}. 

It is well known that over certain frequencies, typical metals can reflect electromagnetic (EM) fields and can thus be used as electromagnetic shielding materials.  In response to the incident EM field, free electrons in the metal collectively oscillate and generate fields of opposite phase relative to the incident field, leading to an exponential decay of the total fields inside the metal.  These evanescent electric ($\bm E$) and magnetic ($\bm H$) fields cannot propagate due to the negative sign of $\epsilon$ and positive $\mu$, intrinsic to metals at the appropriate bandwidth.  The skin depth, i.e., the distance EM fields drop to $1/e$ of their surface value, is determined primarily from the frequency of the source, and the corresponding material parameters.  If the situation arises whereby a structure possess simultaneously negative effective $\epsilon$ and $\mu$, a type of metamaterial referred to as left-handed material (LHM), the EM fields can then propagate inside the material with a negative refraction at the interface, and ideally the system exhibits transparency.  Thus, conventional {\it linear} LHMs cannot be used to shield electromagnetic fields. 

In this letter, we demonstrate that this picture is drastically modified when nonlinearity of the magnetic response is taken into account, creating a controllable shielding effect in LHMs, accompanied by a parametric reflection.  Unlike metals, that behave as passive reflectors or mirrors, a nonlinear metamaterial slab can function as an active mirror that changes the incident field frequency upon reflection.  We show that the nonlinear-induced skin depth, $\delta$, which is a measure of the shielding strength in LHMs, is tunable and much less than the wavelength, $\lambda$, of the EM fields being shielded, even in the absence of material absorption.  It is vastly different in conventional metals, where the skin depth is typically fixed and on the order of $\lambda$.  This shielding effect is the consequence of backward parametric interaction due to the nonlinear magnetic response of the LHM and near-zero refractive index at the pump frequency.  Recently, many interesting phenomena associated with near-zero refractive index metamaterials \cite{Enoch,Husakou,Silve,Ziolkowski} have been reported, such as directive emission\cite{Enoch} and steplike transmission\cite{Husakou}.  In the more conventional materials (positive $\epsilon$ and $\mu$), also called right-handed materials (RHMs), most nonlinear effects originate from the electronic response.  With left-handed materials however, the magnetic permeability is on equal footing and plays a crucial role.  A nonlinear magnetic response can indeed arise in LHMs \cite{Zharov,Lapine,Gorkunov}, and exact phase matching can be achieved between forward and backward waves \cite{Shadrivov,Popov}, leading to enhanced second-harmonic generation\cite{Klein,Mattiucci}.  In the majority of past works, the nonlinear response was restricted to be either electric or magnetic.  Moreover, to simplify the problem, the slowly varying envelope approximation (SVEA) was often used.  In this paper, without recourse to SVEA, a general theory is developed for parametric interactions in metamaterial multilayer structures simultaneously possessing nonlinear electronic and magnetic responses.  By solving Maxwell's equations exactly, multiple reflections at the RHM/LHM interfaces are accounted for, thus properly describing the nonlinear parametric interaction between forward and backward waves.

Consider a 1D $N$-layer periodic structure with period $d=d_r+d_l$, where $d_r$ and $d_l$ are, respectively, the layer thickness of 
the RHM and the LHM.  The total length of the structure is $L$.  The $z$-axis coincides with the direction normal to the plane of the layers.  The structure is configured such that the right-handed medium has a nonlinear electronic response, whereas the left-handed medium has a nonlinear magnetic response.  The linear permittivity in the RHM, $\epsilon_r$, is given by the standard Drude-Lorentz form,
\beq
\label{epsr}
\epsilon_r(\omega) = 1+ \frac{\omega_{pr}^2}{\omega_{r}^2-\omega^2-i\gamma\omega}\,,
\eeq
where $\omega_{pr}$ is the effective plasma frequency, $\omega_r$ is a resonant frequency, and $\gamma$ is the damping factor of the medium.  The linear permeability in the RHM, $\mu_r$, is constant.  In the LHM, the linear permittivity, $\epsilon_l$, and permeability, $\mu_l$, are similarly described by,
\beq
\label{epsl}
\epsilon_l(\omega) = 1+\frac{\omega_{pl}^2}{\omega_{l}^2-\omega^2-i\gamma\omega}, 
\mu_l(\omega) = 1 + \frac{F\omega^2}{\omega_0^2-\omega^2-i\gamma\omega},
\eeq
where $\omega_{pl}$ is the effective plasma frequency, $\omega_l$ and $\omega_0$ are the resonant frequencies of the medium, and $F$ is a geometrical filling factor.  The parametric interaction is a three-wave mixing process in a nonlinear medium that involves a strong pump beam and two relatively weak beams, the signal and idler.  The distinction between signal and idler is arbitrary, just as long as energy and momentum are conserved.  In general, due to its dispersive nature, LHM possess left-handed properties only in certain frequency regimes.  We assume the signal ($\omega_1$) is a negative-index wave, while the idler ($\omega_2<\omega_1$) and the pump ($\omega_3=\omega_1+\omega_2$) are both positive-index waves.  We consider TEM modes propagating along the $z$-direction as an $x$-polarized $\bm E$ field and $y$-polarized $\bm H$ field.  In the RHM, the electric field of the signal, $E_1$, and idler, $E_2$, are thus described by
\begin{equation}
\begin{split}
\label{dE}
\frac{d^2E_1}{dz^2} + \frac{\omega_1^2}{c^2}\epsilon_{1r}\mu_{1r} E_1  &=  -\kappa_{1r} E_2^*  \\
\frac{d^2E_2}{dz^2} + \frac{\omega_2^2}{c^2}\epsilon_{2r}\mu_{2r} E_2  &=  -\kappa_{2r} E_1^* \,.
\end{split}
\end{equation}
Similarly, the corresponding magnetic fields in the LHM are given by,
\begin{equation}
\begin{split}
\label{dH}
\frac{d^2H_1}{dz^2} + \frac{\omega_1^2}{c^2}\epsilon_{1l}\mu_{1l} H_1  &=  -\kappa_{1l} H_2^*  \\
\frac{d^2H_2}{dz^2} + \frac{\omega_2^2}{c^2}\epsilon_{2l}\mu_{2l} H_2  &=  -\kappa_{2l} H_1^* \,,
\end{split}
\end{equation}
where $\kappa_{ir} = (\omega_i/c)^2 (8\pi\mu_{ir}\chi_e^{(2)}E_3)$ and
$\kappa_{il} = (\omega_i/c)^2 (8\pi\epsilon_{il}\chi_m^{(2)}H_3)$, for $i=1,2$, are nonlinear coupling parameters that couple the signal and idler waves.  The coefficients, $\chi_e^{(2)}$ and $\chi_m^{(2)}$ are the second-order nonlinear electric and magnetic susceptibilities, respectively.  Here $E_3$ and $H_3$ represent the pump wave.  With the assumption of plane waves and negligible pump depletion, Eqs.~(\ref{dE}) and (\ref{dH}) can be solved analytically.  In particular, we consider metamaterials with near-zero refractive-index at the pump frequency.  This approach stems from the fact that the pump photons carry negligible momentum in three-wave mixing processes, and thus forward and backward coupling between signal and idler become dominated in the LHM, leading to {\it enhanced} parametric reflection.  Although the refractive index is close to zero, the pump can still be coupled into the structure through matched impedance methods or tailoring the waveguide appropriately\cite{Silve,Ziolkowski}.  

The fields inside each layer are linear superpositions of forward and backward waves.  Each forward and backward wave contains two modes due to the parametric interaction between the signal and idler.  The general EM solutions in the $n$th layer can be succinctly written in terms of the modal fields, $\Phi_{q\beta}^{(n)}(z)$ (for $q=r,l$):
\begin{equation}
\begin{split}
\label{genE}
E_1^{(n)}(z)   &=  \Phi_{r\beta}^{(n)}(z) + C_{1r}\Phi_{r\eta}^{*(n)}(z), \\
E_2^{*(n)}(z) &=  C_{2r}^*\Phi_{r\beta}^{(n)}(z) + \Phi_{r\eta}^{*(n)}(z),
\end{split}
\end{equation}
in the RHM.  For the LHM, we have
\begin{equation}
\begin{split}
\label{genH}
H_1^{(n)}(z)   &=  \Phi_{l\beta}^{(n)}(z) + C_{1l}\Phi_{l\eta}^{*(n)}(z), \\
H_2^{*(n)}(z) &=  C_{2l}^*\Phi_{l\beta}^{(n)}(z) + \Phi_{l\eta}^{*(n)}(z),
\end{split}
\end{equation}
The coupling coefficients $C_{1q}$ and $C_{2q}$ ($q=r,l$) are given by
$C_{1q} \equiv \kappa_{1q}/(\eta_q^{*2}-k_{1q}^2)$, and
$C_{2q} \equiv \kappa_{2q}/(\beta_q^{*2}-k_{2q}^2)$,
where $k_{ir}^2=(\omega_i/c)^2\epsilon_{ir}\mu_{ir}$ and $k_{il}^2=(\omega_i/c)^2\epsilon_{il}\mu_{il}$, for  $i=1,2$.
The modal fields are written in terms of plane waves,
\begin{equation}
\begin{split}
\label{Phi}
\Phi_{q\beta}^{(n)}(z) =& a_{qf\beta}^{(n)}\exp(i\beta_q z) + a_{qb\beta}^{(n)}\exp(-i\beta_q z), \\
\Phi_{q\eta}^{(n)}(z) =& a_{qf\eta}^{(n)}\exp(i\eta_q z) + a_{qb\eta}^{(n)}\exp(-i\eta_q z)\,.
\end{split}
\end{equation}
Inserting Eqs.~(\ref{genE}) and (\ref{genH}) into the nonlinear wave equation yields the propagation constants $\beta_q$ and $\eta_q$,
\begin{equation}
\begin{split}
\label{beta}
\beta_q^2 &= \frac{k_{1q}^2+k_{2q}^{*2}}{2} + \frac{k_{1q}^2-k_{2q}^{*2}}{2}
		\sqrt{1+\frac{4\kappa_{1q}\kappa_{2q}^*}{\left(k_{1q}^2-k_{2q}^{*2}\right)^2}},  \\
\eta_q^2 &= \frac{k_{1q}^{*2}+k_{2q}^2}{2} - \frac{k_{1q}^{*2}-k_{2q}^2}{2}
		\sqrt{1+\frac{4\kappa_{1q}^*\kappa_{2q}}{\left(k_{1q}^{*2}-k_{2q}^2\right)^2}}.
\end{split}
\end{equation}
When the nonlinear parameters $\kappa_{1q}$ and $\kappa_{2q}$ vanish, the propagation constants $\beta_q$ and $\eta_q$ 
correctly reduce to their linear counterparts for the signal and idler, i.e. $\beta_q=k_{1q}$ (signal) and $\eta_q=k_{2q}$ (idler).
Furthermore, the solutions in Eqs.~(\ref{genE}) and (\ref{genH}) reduce to linear signal and idler waves.  When the nonlinear parameters are non-zero, Eqs.~(\ref{genE}) and (\ref{genH}) contain all possible parametric interactions among forward and backward waves between signal and idler.  The modal coefficients $a_{qf\beta}^{(n)}$, $a_{qb\beta}^{(n)}$, $a_{qf\eta}^{(n)}$, and $a_{qb\eta}^{(n)}$ in Eq.~(\ref{Phi}) can be found by matching boundary conditions that require the continuity of tangential $\bm{E}$ and $\bm{H}$ at each interface.  The inherent nonlinear boundary conditions can be avoided using Maxwell's equations:
$H = c/(i\omega\mu_r) \partial_z E$ in the RHM, and 
$E = c/(i\omega\epsilon_l) \partial_z H$ in the LHM.
To proceed, we define state vectors in the RHM and LHM: 
$\ket{ {\bm a_r}^{(n)}} \equiv(a_{rf\beta}^{(n)},a_{rb\beta}^{(n)},a_{rb\eta}^{*(n)},a_{rf\eta}^{*(n)})^T$, and
$\ket {\bm{a_l}^{(n)} }\equiv (a_{lf\beta}^{(n)},a_{lb\beta}^{(n)},a_{lb\eta}^{*(n)},a_{lf\eta}^{*(n)})^T$.
The transformation between state vectors occurs via:
$\ket{ {\bm a_l}^{(n)} } = \bm{T_r} \ket{ \bm{a_r}^{(n)} } $, and
$\ket { {\bm{a_r}}^{(n+1)} } = \bm{T_l} \ket{ {\bm{a_l}}^{(n)} }$, yielding
$\ket { {\bm{a_r}}^{(n+1)} } = \bm{T_lT_r} \ket{ {\bm{a_r}}^{(n)} }$.
The transfer matrices $\bm{T_r}$ and $\bm{T_l}$ are calculated using the Maxwell relations above.  After some tedious algebra, we have, $\bm{T_r}=\bm{I_l}^{-1}\bm{P_r}$ and $\bm{T_l}=\bm{I_r}^{-1}\bm{P_l}$, where,
\beq
\label{Pr}
\bm{P_r} = \begin{pmatrix} \phi_r & \phi_r^{-1} & C_{1r}\theta_r & C_{1r}\theta_r^{-1}  \\
\displaystyle{\frac{\phi_r}{{\cal Z}_{1r\beta}}} & \displaystyle{-\frac{\phi_r^{-1}}{{\cal Z}_{1r\beta}}} &
\displaystyle{\frac{C_{1r}\theta_r}{{\cal Z}_{1r\eta}}} & \displaystyle{-\frac{C_{1r}\theta_r^{-1}}{{\cal Z}_{1r\eta}}}  \\
C_{2r}^*\phi_r & C_{2r}^*\phi_r^{-1} & \theta_r & \theta_r^{-1}  \\
\displaystyle{\frac{C_{2r}^*\phi_r}{{\cal Z}_{2r\beta}^*}} & \displaystyle{-\frac{C_{2r}^*\phi_r^{-1}}{{\cal Z}_{2r\beta}^*}} &
\displaystyle{\frac{\theta_r}{{\cal Z}_{2r\eta}^*}} & \displaystyle{-\frac{\theta_r^{-1}}{{\cal Z}_{2r\eta}^*}}
\end{pmatrix},
\eeq
and,
\beq
\label{Il}
\bm{I_l} = \begin{pmatrix}
{\cal Z}_{1l\beta} & -{\cal Z}_{1l\beta} & {\cal Z}_{1l\eta}C_{1l} & -{\cal Z}_{1l\eta}C_{1l} \\  
1 & 1 & C_{1l} & C_{1l}\\  
{\cal Z}_{2l\beta}^*C_{2l}^* & -{\cal Z}_{2l\beta}^*C_{2l}^* & {\cal Z}_{2l\eta}^* & -{\cal Z}_{2l\eta}^* 
\\  C_{2l}^* & C_{2l}^* & 1 & 1
\end{pmatrix},
\eeq
where,
$\phi_q\equiv\exp(i\beta_qd_q)$, and $\theta_q\equiv\exp(i\eta_q^*d_q)$.  The generalized impedances in the right-handed and left-handed media at the signal ($n=1$) and idler ($n=2$) frequencies are written as
${\cal Z}_{nr\beta} \equiv (\omega_n\mu_{nr})/(c\beta_{nr})$,
${\cal Z}_{nr\eta} \equiv (\omega_n\mu_{nr})/(c\eta_{nr})$, in RHM,
${\cal Z}_{nl\beta} \equiv (c\beta_{nl})/(\omega_n\epsilon_{nl})$,
${\cal Z}_{nl\eta} \equiv (c\eta_{nl})/(\omega_n\epsilon_{nl})$, in LHM. 
Moreover, $\beta_{1q}=\beta_q$, $\beta_{2q}=\beta_q^*$, $\eta_{1q}=\eta_q^*$, and $\eta_{2q}=\eta_q$, $(q=r,l)$.  Similarly,
\beq
\label{Pl}
\bm{P_l} = \begin{pmatrix} {\cal Z}_{1l\beta}\phi_l & \displaystyle{ -\frac{{\cal Z}_{1l\beta}}{\phi_l} }& 
{\cal Z}_{1l\eta}C_{1l}\theta_l & \displaystyle{-\frac{{\cal Z}_{1l\eta}C_{1l}}{\theta_l} }\\  
\phi_l & \phi_l^{-1} & C_{1l}\theta_l & C_{1l}\theta_l^{-1}  \\
{\cal Z}_{2l\beta}^*C_{2l}^*\phi_l & \displaystyle{-\frac{{\cal Z}_{2l\beta}^*C_{2l}^*}{\phi_l} }& 
{\cal Z}_{2l\eta}^*\theta_l & -{\cal Z}_{2l\eta}^*\theta_l^{-1}  \\
C_{2l}^*\phi_l & C_{2l}^*\phi_l^{-1} & \theta_l & \theta_l^{-1}
\end{pmatrix},
\eeq
and,
\beq
\label{Ir}
\bm{I_r} = \begin{pmatrix} 1 & 1 & C_{1r} & C_{1r}  \\
\displaystyle{{\cal Z}^{-1}_{1r\beta}} & \displaystyle{-{\cal Z}^{-1}_{1r\beta}} & 
\displaystyle{C_{1r}{\cal Z}^{-1}_{1r\eta}} & \displaystyle{-C_{1r}{\cal Z}^{-1}_{1r\eta}}  \\
C_{2r}^* & C_{2r}^* & 1 & 1  \\
\displaystyle{C_{2r}^*{\cal Z}_{2r\beta}^{*\,-1}} & \displaystyle{-C_{2r}^*{\cal Z}_{2r\beta}^{*\, -1}} &
\displaystyle{{\cal Z}_{2r\eta}^{*\,-1}} & \displaystyle{-{\cal Z}_{2r\eta}^{*\,-1}}
\end{pmatrix}.
\eeq

\begin{figure}
\centering
\includegraphics[width=.4\textwidth]{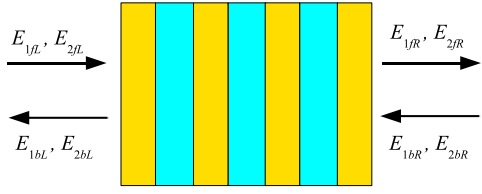}
\caption{(Color online) Metamaterial multilayer considered in this paper.  The amplitudes of the left input fields are represented by $E_{1fL}$ and $E_{2fL}$, and the right ones by $E_{1bR}$ and $E_{2bR}$.  The amplitudes of the left output fields are represented by $E_{1bL}$ and $E_{2bL}$, and by $E_{1fR}$ and $E_{2fR}$ for the right side of the stack.  The subscripts 1 and 2 refer to the signal and idler waves, respectively.}
\label{Fig2_Setup}
\end{figure}

The input and output layers are arranged to be RHM, and the signal $(\omega_1)$ and idler $(\omega_2)$ waves can be incident from both the right and left sides of the structure, as shown in Fig.~\ref{Fig2_Setup}.  By matching tangential fields at the input and output interfaces, the transmission and reflection can be calculated via,
$\ket {\bm{U} } = \bm{A}^{-1}(\ket{\bm{S_R}}-\bm{M}\ket{\bm{S_L}})$,
where $\ket{\bm{U}} \equiv (E_{1fR}, E_{2fR}^*, E_{1bL}, E_{2bL}^*)^T$.
The left source, $\bm{S_L}$, is written as
$\ket{\bm{S_L}} \equiv (E_{1fL}, E_{1fL}{\cal Z}^{-1}_{1L}, E_{2fL}^*, E_{2fL}^*{\cal Z}^{*\,-1}_{2L})^T$, and
the right source, $\bm{S_R}$, compactly as
$\ket{\bm{S_R}} \equiv (E_{1bR}, -E_{1bR}{\cal Z}^{-1}_{1R}, E_{2bR}^*, -E_{2bR}^*{\cal Z}_{2R}^{*\, -1})^T$,
where the ambient impedances at the frequencies $\omega_1$ and $\omega_2$ for the right and left sides of the structure are,
${\cal Z}_{iR}\equiv(\omega_i\mu_{iR})/( ck_{iz})$ and ${\cal Z}_{iL}\equiv (\omega_i\mu_{iL})/(ck_{iz})$.
The $4\times4$ matrix $\bm{A}$ is given by
\beq
\label{AR}
\bm{A}=\begin{pmatrix} -\bm{R_1} & \bm{M_1L_1}+\bm{M_2L_2} \\ -\bm{R_2} & \bm{M_3L_1}+\bm{M_4L_2}\end{pmatrix}\,,
\eeq
where,
\begin{equation}
\begin{split}
\label{RL12}
\bm{R_1}&\equiv\begin{pmatrix} 1 & 0 \\ {\cal Z}_{1R}^{-1} & 0 \end{pmatrix},\qquad
\bm{R_2}\equiv\begin{pmatrix} 0 & 1 \\ 0 & {\cal Z}_{2R}^{*-1} \end{pmatrix}, \\ 
\bm{L_1}&\equiv\begin{pmatrix} 1 & 0 \\ -{\cal Z}_{1L}^{-1} & 0 \end{pmatrix},\qquad
\bm{L_2}\equiv\begin{pmatrix} 0 & 1 \\ 0 & -{\cal Z}_{2L}^{*-1} \end{pmatrix}\,.
\end{split}
\end{equation}
The $2\times2$ submatrices $\bm{M_i}\,\,(i=1,2,3,4)$ arise from the $4\times4$ matrix $\bm{M}$,
\beq
\label{MR}
\bm{M} =\bm{P_rTI_r}^{-1}\equiv\begin{pmatrix} \bm{M_1} & \bm{M_2} \\ \bm{M_3} & \bm{M_4}\end{pmatrix}\,,
\eeq
where $\bm{P_r}$ and $\bm{I_r}$ are given by Eqs.~(\ref{Pr}) and (\ref{Ir}).  The transfer matrix $\bm{T}$ expresses the modal coefficients in the $N$th layer in terms of the first one: $\ket{\bm{a}^{(N)}}\equiv\bm{T}\ket{\bm{a}^{(1)}}$.

\begin{figure}
\centering
\includegraphics[width=.4\textwidth]{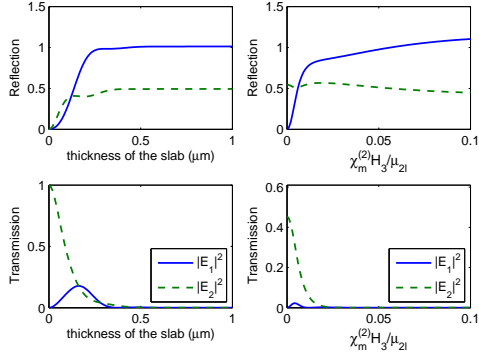}
\caption{(Color online) Transmission and reflection of signal ($E_1$) and idler ($E_2$) for a single nonlinear LHM slab as a function of slab thickness (left panel) and $\chi_m^{(2)}H_3/\mu_{2l}$ (right panel).  The incident field is at the idler wavelength ($\lambda_2=3.6\,\mu$m) and the pump is set to $1.2\,\mu$m, thus the signal is $\lambda_1=1.8\,\mu$m.  At the idler wavelength, $\mu_{2l}=1.72$.  In the left panel, $\chi_m^{(2)}H_3/\mu_{2l}$ is fixed at $0.06$ while in the right panel, the thickness of the slab remains at $1\,\mu$m.}
\label{Fig3_Mirror}
\end{figure}

In the following, we consider the pump wavelength to be operating at $\lambda_3=1.2\,\mu$m (corresponding to $n_3\approx0$),
and take $F=0.77$, $\omega_0=120$\,THz, $\omega_{pl}=300$\,THz, $\omega_{pr}=150$\,THz, $\omega_r=200$\,THz, $\omega_l=120$\,THz, and $\gamma=0$.  We begin by investigating parametric reflection and transmission through a single slab of nonlinear LHM.  Figure~\ref{Fig3_Mirror} shows the transmission and reflection normalized to the incident electric field at the idler wavelength, $\lambda_2=3.6\,\mu$m.  There is no signal frequency at the input; the signal is generated inside the stack due to the nonlinear effect of the materials.  Thus, the signal reflection ($E_1$) exhibited in the plots implies backward transmission.  The frequency  of the signal, $\omega_1$, is determined by the frequencies of the idler and pump via energy conservation, $\omega_1=\omega_3-\omega_2$, yielding $\lambda_1=1.8\,\mu m$.  The left panel shows that the signal and idler reflections increase with slab thickness, while transmission decreases.  The right panel illustrates the variation of the reflection and transmission with the pump field when the thickness of the slab is fixed at $1\,\mu$m.  

\begin{figure}
\centering
\includegraphics[width=.4\textwidth]{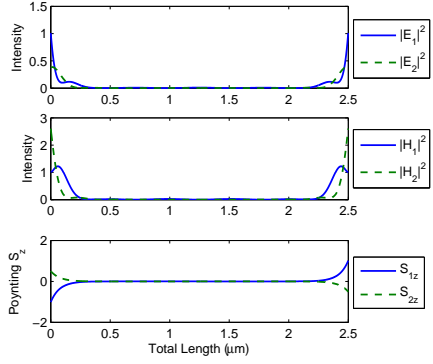}
\caption{(Color online) Parametric shielding by a trilayer metamaterial stack (see text).  The thickness of the LHM and RHM layers are $0.5\,\mu$m and $1.5\,\mu$m, respectively.  The incident field at the idler wavelength ($3.6\,\mu$m) impinges both sides.  Both signal (solid blue) and idler (dashed green) decay exponentially inside the stack.  The skin depth is $\sim0.1\,\mu$m, with  
$\chi_e^{(2)}E_3=0.01\epsilon_{2r}$, $\chi_m^{(2)}H_3=0.06\mu_{2l}$, where $\epsilon_{2r}=1.68$ and $\mu_{2l}=1.72$ at the idler wavelength.}
\label{Fig4_Field_InIdl}
\end{figure}

\begin{figure}
\centering
\includegraphics[width=.35\textwidth]{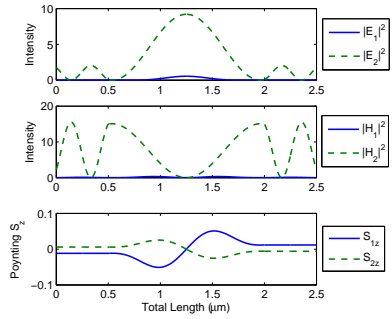}
\caption{(Color online) The results when $\chi_m^{(2)}=0$.  Other parameters and inputs are the same as those in Fig.~\ref{Fig4_Field_InIdl}.  The fields are significant inside the stack.  The absence of EM shielding is clearly evident.}
\label{Fig5_Field_NoNLH}
\end{figure}

To demonstrate electromagnetic shielding, we now consider a three-layer structure composed of RHM sandwiched between two layers of LHM.  An electromagnetic field at the idler wavelength $\lambda_2=3.6\,\mu$m is incident from both sides of the stack.  Shown in Fig.~\ref{Fig4_Field_InIdl} are the intensities of the electric (top panel) and magnetic (middle panel) fields normalized to the input electric field.   The $z$ component of the Poynting vector inside the stack for the signal and idler is also shown in the bottom panel.  As before, there is no signal frequency at the input; the signal is generated inside the stack where the backward parametric interaction between the signal and idler in the negative-index layer leads to an exponential decay from the input interfaces.  After a slight penetration, the electromagnetic fields are virtually zero.  Thus, a nonlinear negative-index layer can be used as a parametric mirror that changes frequency upon reflection via the nonlinear interaction in the penetration layer.  
Remarkably, the skin depth can be 30 times smaller than the wavelength of the incident field, as seen by the approximate expression,
\beq
\label{skin}
\delta\approx \frac{\sqrt{2}}{16\pi^2}\frac{\sqrt{|\mu_{1l}\mu_{2l}|}}{|\chi_m^{(2)}H_3|}
\sqrt{\frac{\lambda_1^2}{|\epsilon_{1l}\mu_{1l}|} + \frac{\lambda_2^2}{|\epsilon_{2l}\mu_{2l}|}}\, ,
\eeq
showing explicitly that $\delta$ is inversely proportional to the pump field strength.  The dramatic effects of the nonlinear shielding vanish if the nonlinear magnetic response of the LHM is zero, as exhibited in Fig.~\ref{Fig5_Field_NoNLH}, where the same quantities as those in Fig.~\ref{Fig4_Field_InIdl} are plotted except with $\chi_m^{(2)}=0$.  The top two panels of Fig.~\ref{Fig5_Field_NoNLH} show that the electromagnetic fields are significant inside the stack.  The bottom panel shows the
Poynting vector, where the signal generated in the nonlinear RHM (the middle section from $0.5\,\mu$m to $2\,\mu$m) propagates straight through the linear LHM (the two end sections) without idler interaction.  The small but finite energy flow arises from the nonlinear effect in the RHM.  For a purely linear structure, a signal wave cannot be generated, and the idler becomes a standing wave with net zero power flow inside the stack due to oppositely directed input fields with identical amplitudes. 

This work is supported by NAVAIR's ILIR program sponsored by ONR.

\end{document}